\documentclass[10pt]{article}
\usepackage{epsfig}
\usepackage{notconf}
\begin{document}


\title{Finding typical high redshift galaxies with the NOT}
\author{J.U. Fynbo and P. M\o ller\\
    European Southern Observatory, Karl-Schwarzschild-Stra\ss e 2,\\
    D-85748 Garching-bei-M\"unchen, Germany\\
    B. Thomsen\\
    Institute of Physics and Astronomy, University of \AA rhus,
    D-8000 \AA rhus C, Denmark}

\maketitle

\begin{abstract}
  We present results from an ongoing search for galaxy counterparts of a
  subgroup of Quasar Absorption Line Systems called Damped Ly$\alpha$
  Absorbers (DLAs). DLAs have several characteristics that make them
  prime candidates for being the progenitors of typical present day 
  galaxies.

\end{abstract}

\section{Damped Ly$\alpha$ Absorbers and high redshift galaxies}
Damped Ly$\alpha$ Absorbers are QSO absorption line systems
with HI column density larger than $2\times10^{20} cm^{-2}$.
This very large column density absorption occurs in regions 
of self shielding, cooled gas,
i.e. where we expect stars to form. Hence DLAs are prime
candidates for being the progenitors of present day galaxies.
This hypothesis is strengthened by the fact that the neutral 
gas content of DLAs at high redshift, within the uncertainties, 
is known to be the same as that of visible matter in present 
day galaxies (Wolfe et al., 1995). Hence, DLAs being HI column 
density selected galaxies are truly representative of the 
progenitors of present day galaxies.

   There are primarily two pieces of information we wish to obtain 
via the study of DLAs : {\it (i)} the size and {\it (ii)} the stellar 
content of typical high redshift galaxies. Concerning {\it (i)}, it
is a long standing controversy whether DLAs are large fully formed disk
galaxies or small merging galaxy subunits (e.g. Wolfe et al., 1986, 
Haehnelt et al., 1998).
The actual size of typical high redshift galaxies will give us 
information about the nature of the dark matter that forms the haloes 
containing the baryons. Concerning {\it (ii)}, it has become clear that 
the star formation histories of all local group members differ from 
that of the Milky Way 
and differ amongst each other. Therefore we cannot expect any 
single galaxy in the local group to be a good tracer of the global 
star formation history (e.g. Tolstoy, 1998). Another line of approach 
that has been pursued heavily in the last few years has been to try 
to obtain the global star formation history via the study of so called 
Lyman break galaxies (LBGs) in the early universe. LBGs are found
using a technique based on the fact that young, star forming
galaxies will have a strong spectral break at the lyman limit, which at 
high redshift is redshifted into the optical window (Steidel et al.,
1996). LBGs need to be bright enough for
spectroscopical confirmation of their high redshift so they are typically 
brighter than R(AB)=26. Assuming that DLAs arise in gaseous discs associated 
with LBGs we can compare DLAs and LBGs by calculating how faint we 
need to integrate down the extrapolation of the luminosity function of 
LBGs in order to explain the observed probability for a QSO line of sight 
to cross a DLA. Results of this calculation are presented in Fynbo et 
al., 1999, and summarized here. At $z = 3$ we find that 70-90\% of DLA 
galaxy counterparts are fainter than the current limit for spectroscopic 
confirmation of LBG candidates of R(AB)=26. Hence LBGs are highly 
atypical high redshift galaxies, probably the progenitors of present day
bright cluster galaxies (Baugh et al., 1998). Studying high redshift 
DLAs is therefore the only way to obtain information about the 
nature of {\it typical} (in that they contain the baryons found in
galaxies today) galaxies in the early universe.

The most obvious method by which to determine the sizes and stellar
contents of DLAs is to detect emission from them.

\section{Imaging of DLAs with the NOT}
\begin{figure}[t]
\epsfig{file=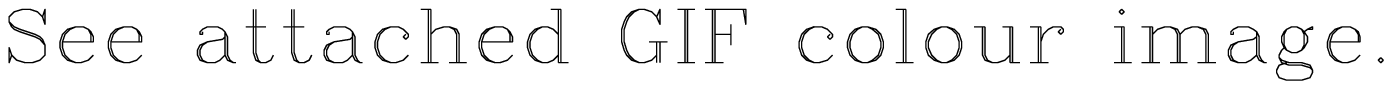, width = 12cm}
\caption{Results from narrow/broad imaging of two DLA fields with the
NOT. {\it Left :} the DLA towards Q0151+048A. {\it Right :} 
the DLA towards PKS1157+014.} 
\label{fig-1}
\end{figure}

   From an observational point of view the main problems in studying
emission from DLAs are {\it (i)} that they are very faint and {\it (ii)} 
the presence of a much brighter QSO at a distance of only 0-3 arcsec  
on the sky. In the spectrum of the background QSOs DLAs at redshifts 
$z\approx2$ produce regions of 15-25\AA \ (the width depending on the 
HI column density) of saturated absorption. Hence imaging in a narrow filter 
with a width corresponding to the width of the damped absorption line 
will circumvent problem {\it (ii)}. If the DLA is a Ly$\alpha$ 
{\it{emitter}} it will be relatively easy to detect against the
modest sky background in the narrow band filter which helps
circumventing problem {\it (i)}. Narrow band imaging of DLAs have been 
pursued in more 
than a decade (e.g. Lowenthal et al., 1995), but only recently with 
success. The DLA at $z = 2.81$ towards the $z_{em}=2.79$ PKS0528-250 
(M\o ller and Warren, 1993, 1998, Warren and M\o ller, 1996) was detected 
with narrow band 
imaging using the ESO 3.6m telescope and confirmed by spectroscopy on 
the ESO NTT. Here we describe our results on narrow and broad band 
imaging of the DLAs towards Q0151+048A (z$_{abs}$=1.9342) and PKS1157+014
(z$_{abs}$=1.9436). These two DLAs were chosen because they are
$z_{abs}\approx z_{em}$ systems, as is the DLA towards PKS0528-250. 
The QSO redshifts are $z_{em}=1.921$ and $z_{em}=1.978$ for Q0151+048A and
PKS1157+014 respectively. Moreover, Q0151+048 is very interesting in
being a physical QSO pair (not a lensed system) with two QSOs at nearly 
the same redshift (Meylan, et al., 1990). The B component has redshift 
$z_{em}=1.937$ (M\o ller, Warren and Fynbo, 1998). Moreover, the
DLA towards PKS1157+014 has one of the highest HI column densities 
($6\times10^{21} cm^{-2}$) of all known DLAs. NOT was 
the perfect instrument for these DLAs due to the high spacial resolution 
and the very high UV sensitivity of the Loral CCD at the wavelength of 
redshifted Ly$\alpha$ ($\approx$3600\AA). 

The DLA towards Q0151+048 was imaged in narrow-band, U and I in four
nights in September
1996 with StanCam. We obtained 5$\sigma$ point source detection limits of
n(3567)=24 (corresponding to $5.0\times10^{-17} erg s^{-1} cm^{-2}$ for
Ly$\alpha$ at the absorption redshift), I(AB)=25.7 and U(AB)=26.0 
respectively. The three left panels in fig.~\ref{fig-1} show 
$100\times60 arcsec^2$ from the combined I-frame (top), U-frame (middle)
and narrow band frame (bottom). North is up and east is to the left. Seen 
are the two QSOs in the center of the frames and a candidate $z=1.93$ 
Ly$\alpha$ emitting galaxy 40$\arcsec$ \ east of the QSOs. In the lower 
frame we have subtracted the Point-Spread-Functions of the two QSOs so
that the extended Ly$\alpha$ emission from the DLA is clearly seen.

The DLA towards PKS1157+014 was imaged in narrow-band, U and I in two
nights in March 1998 with Alfosc. We obtained 5$\sigma$ point source 
detection limits of n(3567)=23.2 (corresponding to 
$7.5\times10^{-17} erg s^{-1} 
cm^{-2}$ for Ly$\alpha$ at the absorption redshift), I(AB)=25.9 and 
U(AB)=25.3 respectively. The three right panels in fig.~\ref{fig-1} show 
extractions from the combined I-frame (top), U-frame (middle)
and narrow band frame (bottom) with the same field size as the left frames,
but with north to the left and east down. Seen are the QSO in the center 
and two candidate $z=1.94$ Ly$\alpha$ emitting galaxies. In the lower
frame the QSO has completely vanished due to the extremely strong damped
absorption line. There is no evidence for Ly$\alpha$ emission from the
DLA at impact parameters smaller than 10$\arcsec$.

\section{Discussion}

\begin{figure}[t]
\begin{center}
\epsfig{file=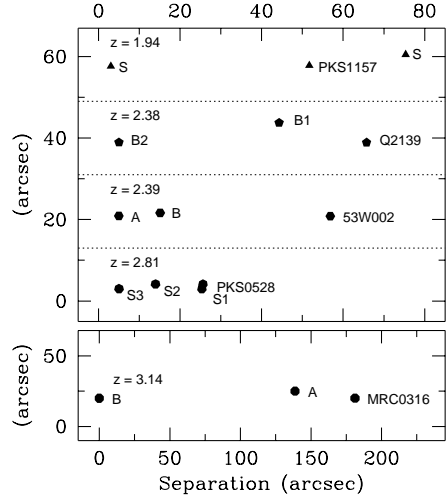, width = 8cm}
\caption{Evidence for alignment in groups of Ly$\alpha$ emitting
galaxies} 
\label{fig-2}
\end{center}
\end{figure}

   It is not yet possible to draw general conclusion about the nature
of DLA galaxies based only on the few DLA galaxy counterparts currently
studied in emission. We do, however, note that in all three DLA fields studied
with narrow band imaging so far we have found one or more candidate
galaxies at the DLA redshift. In the right frames of fig.~\ref{fig-1}
the two galaxies seem to be aligned with the QSO. As noted by
M\o ller and Warren, 1998, there is growing evidence for
filamentary structure in the distribution of Ly$\alpha$ emitting galaxies 
at high redshift. Fig.~\ref{fig-2} is an updated version of their Fig.6 
showing alignments in 5 galaxy groups at high redshift, including the field 
around PKS1157+014. This trend is in agreement with N-body simulations
of hierarchical structure formation were galaxies predominantly form along 
filaments (e.g. Evrard et al., 1994).

\end{document}